\documentclass[12pt]{article}
\usepackage[dvips]{graphics}

\begin{document}

\vspace{1cm}

\begin{center}
\LARGE{Neutrinos in the Electron}

\bigskip
\Large{ E.L. Koschmieder}

\bigskip
\small{ Center for Statistical Mechanics\\
The University of Texas at Austin, Austin TX 78712, USA\\
e-mail: koschmieder@mail.utexas.edu}

\end{center}

\bigskip
\noindent
\small
{We will show that one half of the rest mass of the electron is equal to the sum
 of the rest masses of electron neutrinos and that the other half of the rest mass
of the electron is given by the energy in the sum of electric oscillations. With this 
composition we can explain the rest mass, the electric charge, the spin and
the magnetic moment of the electron.}  

\normalsize

\section*{Introduction}

   After J.J. Thomson [1] discovered the small corpuscle which soon 
 became known as the electron an enormous amount of theoretical work
 has been done to explain the existence of the electron. Some of the 
most distinguished physicists have participated in this effort. Lorentz [2],
 Poincar\'{e} [3], Ehrenfest [4], Einstein [5], Pauli [6], and others showed 
that it is fairly certain that the electron cannot be explained as a purely 
electromagnetic particle. In particular it was not clear how the electrical 
charge could be held together in its small volume because the internal 
parts of the charge repel each other. Poincar\'{e} [7] did not leave it at 
showing that such an electron could not be stable, but suggested a 
solution for the problem by introducing what has become known as 
the Poincar\'{e} stresses whose origin however remained unexplained.
 These studies were concerned with the static properties of the 
electron, its mass m(e$^\pm$) and its electric charge e.
In order to explain the electron with its existing mass and charge
it appears to be necessary to add to Maxwell's equations a 
non-electromagnetic mass and a non-electromagnetic force which 
could hold the electric charge together.
 We shall see what this mass and force is. 

   The discovery of the spin of the electron by Uhlenbeck and 
Goudsmit [8] increased the difficulties of the problem in so far as it now
 had also to be explained how the angular momentum
 $\hbar$/2 and the magnetic moment $ \mu_e$ come about.
The spin of a point-like electron seemed to be explained by Dirac's [9] 
equation, however it turned out later [10] that Dirac type equations can be 
constructed for any value of the spin. Afterwards Schr\"{o}dinger [11] 
tried to explain the spin and the magnetic moment of the electron with 
 the so-called Zitterbewegung. Later on many other models of the electron
 were proposed. On p.74 of his book ``The Enigmatic Electron" Mac Gregor 
[12] lists more than thirty such models.
At the end none of these models has been completely successful 
because the problem developed a seemingly insurmountable difficulty when 
it was shown through electron-electron scattering experiments that the radius 
of the electron must be smaller than $10^{-16}$\,cm, in other words that the
 electron appears to be a point particle, at least by three orders of 
magnitude smaller than the classical electron radius r$_e$ = e$^2$/mc$^2$ = 2.8179$\cdot10^{-13}$\,cm. This, of 
course, makes it very difficult to explain how a particle can have
a finite angular momentum when the radius goes to zero, and how an 
electric charge can be confined in an infinitesimally small volume. If the 
elementary electrical
charge were contained in a volume with a radius of O($10^{-16}$)\,cm the
 Coulomb self-energy would be orders of magnitude larger than the rest 
mass of the electron, which is not realistic. The choice is between a 
massless point charge and a finite size particle with a non-interacting mass 
to which an elementary electrical charge is attached.

   We propose in the following that the non-electromagnetic mass which 
seems  to be necessary in order to explain the mass of the electron 
consists of neutrinos. This is actually a necessary consequence of our 
standing wave model [13] of the masses of the mesons and baryons. 
And we propose that the non-electromagnetic force required to 
hold the  electric charge and the neutrinos in the electron together is the 
weak nuclear force which, as we
 have suggested in [13], holds together the masses of the mesons
 and baryons and also the mass of the muons. Since the range of the weak 
nuclear force is on the order of $10^{-16}$\,cm the neutrinos can only be 
arranged in a lattice with the weak force extending from each lattice point only
to the nearest neighbors.  The size of the neutrino lattice in the electron does 
not at all contradict the results of the scattering experiments, just as the 
explanation of the mass of the  muons with the standing wave model 
 does not contradict the apparent point particle characteristics of the muon,
 because neutrinos are in a very good approximation non-interacting and 
therefore are not noticed in scattering experiments with electrons. 

\section{ The mass and charge of the electron}

   The rest mass of the electron is m(e$^\pm$) = 0.510\,998\,92 $\pm$ 
4$\cdot10^{-8}$\,MeV/c$^2$ and the electrostatic charge
 of the electron is e = 4.803\,204\,41$\cdot10^{-10}$\,esu, as stated 
in the Review  of Particle Physics [14]. Both are 
known with great accuracy. The objective of a theory of the 
electron must be the explanation of both values. We will first explain
 the rest mass of the electron making use of what we
 have learned from the standing wave model, in particular of what
 we have learned about the explanation of the mass of the 
$\mu^\pm$\,mesons in [13]. 
The muons are leptons, just as the electrons, that means that they interact 
with other particles exclusively through the electric force. The muons 
have a mass which is 206.768 
times larger than the mass of the electron, but they have the same 
elementary electric charge as the electron or positron and the same 
spin. Scattering experiments tell that the $\mu^\pm$\,mesons are point 
particles  with a size $<$\,$10^{-16}$\,cm, just as the electron. In other 
words, the muons have the same characteristics as the electrons and
 positrons but for a mass which is about 200 times larger. Consequently 
the muon is often referred to as a ``heavy"  electron. If a 
non-electromagnetic  mass is required to explain
 the mass of the electron then a non-electromagnetic mass 200 times
 as large as in the electron is required to explain the mass of the muons.
These non-electromagnetic masses must be \emph{non-interacting}, 
otherwise scattering experiments could not find the size of either the 
electron or the muon at 10$^{-16}$\,cm. 
 
   We have already explained the mass of the muons with the standing wave 
model [13]. According to this model the muons consist of an elementary 
electric charge and a lattice of neutrinos  
 which, as we know, do not interact with charge or mass.  Neutrinos 
are the only non-interacting matter we know  of. 
In the muon lattice  are, according to [13],
 (N\,-\,1)/4 = N$^\prime$/4 muon neutrinos $\nu_\mu$  (respectively anti-muon
  neutrinos $\bar{\nu}_\mu$), N$^\prime$/4 electron neutrinos $\nu_e$ 
and the same number of anti-electron neutrinos $\bar{\nu}_e$,  one 
elementary electric charge and the energy of the lattice oscillations. 
The letter N stands for the number of all neutrinos and antineutrinos in
 the cubic lattice of the $\pi^\pm$ mesons [13,\,Eq.(15)]

\begin{equation} \mathrm{N} = 2.854\cdot10^9\,. \end{equation}
It is, according to [13], a necessary consequence of the decay of the
 $\mu^-$ muon $\mu^- \rightarrow$ e$^- + \bar{\nu}_e + \nu_\mu$ that there
 must be N$^\prime$/4 electron neutrinos $\nu_e$ in the emitted electron,
where N$^\prime$ = N - 1 $\cong$ N [13]. 
For the mass of the electron neutrinos and anti-electron neutrinos we found
in Eq.(34) of [13] that

\begin{equation} \mathrm{m}(\nu_e) = \mathrm{m}(\bar{\nu}_e) =
 0.365 \,\mathrm{milli\,eV/c^2}\,. \end{equation}

\noindent
The sum of the energies in the rest masses of the N$^\prime$/4 neutrinos
or antineutrinos  in the lattice of the electron or positron is then

\begin{equation} \sum{\,\mathrm{m(\nu_e)c^2}} =
 \mathrm{N}^\prime/4\cdot\mathrm{m}(\nu_e)\mathrm{c}^2 = 
0.260\,43\,\mathrm{MeV}
 = 0.5096\,\mathrm{m(e^\pm)}\mathrm{c}^2\,. \end{equation}

   To put this in other words, one half of the
 rest mass of the electron comes from the rest masses of
 electron neutrinos. The other half of the rest mass of the
 electron must originate from the energy in the electric charge carried  
by the electron. From pair production 
$\gamma$ + M   $\rightarrow$  e$^-$ + e$^+$ + M, (M being any
 nucleus), and from conservation of neutrino numbers follows necessarily 
that there must also be a neutrino lattice 
composed of N$^\prime$/4 anti-electron neutrinos, which make up 
the lattice of the positrons, which lattice has, because of Eq.(2), the 
same rest mass as the neutrino lattice of the electron, as it must be for  
the antiparticle of the electron.

   Fourier analysis dictates that a continuum of high frequencies must be in 
the electrons or positrons created by pair production in a timespan of 
$10^{-23}$ seconds. We will now determine the energy E$_\nu$(e$^\pm$)
contained in the oscillations in the  interior of the electron. Since
 we want to explain the \emph{rest mass} of the electron we can only 
consider the frequencies of non-progressive waves, either standing waves
or circular waves. The sum of the energies of the lattice
oscillations is, in the case of the $\pi^\pm$\,mesons,  given by 

\begin{equation} \mathrm{E}_\nu(\pi^\pm) =  
\frac{\mathrm{h}\nu_0\mathrm{N}}{2\pi(\mathrm{e^{h\nu/kT}}\,\mathrm{-}\,1)}
\,\int\limits_{-\pi}^{\pi}\,\phi\,d\phi\,. \end{equation}
\noindent
This is Eq.(14) combined with Eq.(16) in [13] where they were used to 
determine the oscillation energy
 in the $\pi^0$ and $\pi^\pm$ mesons. This equation was introduced 
by Born and v.\,Karman [15] in order to explain the internal 
energy of cubic crystals. In Eq.(4) h is Planck's constant,
 $\nu_0$  = c/2$\pi\emph{a}$ is the reference frequency with the lattice 
constant \emph{a} = $10^{-16}$ cm, N is the number of all oscillations,
 $\phi = 2\pi\emph{a}/\lambda$ and T is the temperature in the lattice, 
for which we found in [13] the value T = 2.38\,$\cdot$\,$10^{14}$\,K. 
If we apply Eq.(4) to  
the oscillations  in the electron which has N$^\prime$/4 electron neutrinos
 $\nu_e$ we arrive at  E$_\nu$(e$^\pm)$  = 1/4$\cdot$E$_\nu(\pi^\pm$), 
which is mistaken because E$_\nu(\pi^\pm$) $\approx$ m($\pi^\pm$)c$^2$/2
and m($\pi^\pm$) $\approx$ 273\,m(e$^\pm$). Eq.(4) must
 be modified in order to be suitable for the oscillations in the electron. 
It turns out that we must use
\begin{equation} \mathrm{E}_\nu(\mathrm{e}^\pm) =  
\frac{\mathrm{h}\nu_0\mathrm{N}\cdot\alpha_f}{2\pi(\mathrm{e^{h\nu/kT}}\,
\mathrm{-}\,1)}\,\int\limits_{-\pi}^{\pi}\,\phi\,d\phi\,,
\end{equation}
\noindent
where $\alpha_f$ is the fine structure constant. The appearance of 
$\alpha_f$ in Eq.(5)
indicates that the nature of the oscillations in the electron is different 
from the oscillations 
in the $\pi^0$ or $\pi^\pm$ lattices. With $\alpha_f$ = e$^2/\hbar$c and 
$\nu_0$ = c/2$\pi$\emph{a} we have 
\begin{equation}h\nu_0\alpha_f = e^2/\emph{a}\, \end{equation} 
that means that the oscillations in the electron are \emph{electric oscillations}.

   There must be N$^\prime$/2 oscillations of the elements of the electric 
charge  in e$^\pm$, because we deal with non-progressive waves, the 
superposition of two waves. As we will see later the spin requires 
that the oscillations are circular. That means that  2$\times$N$^\prime$/4 
$\cong$ N/2 oscillations are in Eq.(5). From Eqs.(4,5) then follows that

\begin{equation} \mathrm{E_\nu(e}^\pm) =
 \alpha_f/2\cdot\mathrm{E}_\nu(\pi^\pm)\,. \end{equation}
\noindent
E$_\nu(\pi^\pm)$  is the oscillation energy in the $\pi^\pm$\,mesons which  
can be calculated with Eq.(4). According to Eq.(27) of [13] it is
\begin{equation} \mathrm{E}_\nu(\pi^\pm) = 67.82 \,\mathrm{MeV} =
0.486\,\mathrm{m}(\pi^\pm)\mathrm{c}^2 \approx 
\mathrm{m}(\pi^\pm)\mathrm{c}^2/2\,.
 \end{equation}
 With E$_\nu(\pi^\pm$) $\approx$ m($\pi^\pm$)c$^2$/2 = 139.57/2\,MeV 
and $\alpha_f$  = 1/137.036  follows from Eq.(7) that 
\begin{equation} \mathrm{E_\nu(e}^\pm) = \frac{\alpha_f}{2}\cdot
\frac{\mathrm{m}(\pi^\pm)\mathrm{c}^2}{2} = 0.254\,62\,\mathrm{MeV}
= 0.99657\,\mathrm{m(e^\pm)}\mathrm{c}^2/2\,. 
\end{equation}
\noindent
We have determined the value of the oscillation energy in e$^\pm$ from 
the product of the very accurately known fine structure constant and the 
very accurately known rest mass of the $\pi^\pm$\,mesons. \emph{One half 
of the energy in the rest mass of the electron comes from the electric oscillations 
in the electron}. The other half of the energy in the rest mass of the electron is
 in the rest masses of the neutrinos in the electron.

  We can confirm Eq.(9) 
using Eq.(5) or Eq.(13) with N/2 = 1.427$\cdot10^9$, 
e = 4.803$\cdot10^{-10}$\,esu,
 $\emph{a}$ = 1$\cdot10^{-16}$\,cm, f(T) = 1/1.305$\cdot10^{13}$, and 
with the integral being $\pi^2$ we obtain E$_\nu$(e$^\pm$) = 
0.968\,m($\mathrm{e}^\pm)$c$^2$/2.
This calculation involves more parameters than Eq.(9) and is consequently
less accurate than Eq.(9).

   In a good approximation the oscillation energy of e$^\pm$ in Eq.(9) is
 equal to the sum of the energies in the rest masses
 of the electron neutrinos in the e$^\pm$ lattice in Eq.(3). Since
 \begin{equation} \mathrm{m(e}^\pm)\mathrm{c}^2
 =  \mathrm{E}_\nu(\mathrm{e}^\pm) + \sum{\,\mathrm{m}(\nu_e)\mathrm{c}^2} 
= \mathrm{E_\nu(e^\pm)} + \mathrm{N^\prime/4\cdot m(\nu_e)c^2}\,,
\end{equation}
\noindent 
it follows from Eqs.(3) and (9) that  
\begin{equation} \mathrm{m(e^\pm)c^2(theor)} = 0.5151\,\mathrm{MeV} = 
1.0079\,\mathrm{m(e^\pm)c^2(exp)}\,. \end{equation}
The measured rest mass of the electron or positron agrees within the accuracy
 of the parameters N and m($\nu_e)$ with the theoretically 
predicted rest masses.

   From Eq.(7) follows with E$_\nu(\pi^\pm)$ $\cong$  m($\pi^\pm$)c$^2$/2  that

\vspace{0.5cm} 
\centerline{2E$_\nu$(e$^\pm)$ $\cong$ m(e$^\pm$)c$^2$
= $\alpha_f$E$_\nu(\pi^\pm)$ = $\alpha_f$m$(\pi^\pm)$c$^2$/2\,,}

\vspace{0.5cm}
\noindent
or that
\begin{equation} \mathrm{m(e^\pm)}\cdot2/\alpha_f  = 
274.072\,\mathrm{m(e^\pm)} 
 \cong \mathrm{m(\pi^\pm)}\,, \end{equation}
whereas the actual ratio of the mass of the $\pi^\pm$\,mesons  to the mass
of the electron is 
m($\pi^\pm$)/m(e$^\pm$) = 273.132 or 0.9965$\cdot$2/$\alpha_f$. We
have here recovered the ratio m($\pi^\pm$)/m(e$^\pm$) which we found
with the standing wave model of the $\pi^\pm$\,mesons, Eq.(65) of [13].
This seems to be a necessary condition for the validity of our model of
the electron.

   We have thus shown that the \emph{rest mass of the electron can be 
explained} by the sum  of the rest masses of the electron neutrinos in a cubic 
 lattice with N$^\prime$/4 electron neutrinos $\nu_e$ and the mass in the 
sum of the energy of N/2 electric oscillations in the lattice, Eq.(9). The one 
oscillation added to the 2$\times$N$^\prime$/4 oscillations is the 
oscillation at the center of the lattice, Fig.(1). From this model follows, since 
it deals with a cubic neutrino lattice, that \emph{the electron is not a point
particle}, which is unlikely to begin with, because at a true point the self-energy 
would be infinite. However, since neutrinos are non-interacting their presence 
will not be detected in electron-electron scattering experiments. 

\begin{figure}[h]
\vspace{0.5cm}
\hspace{2.2cm}
\includegraphics{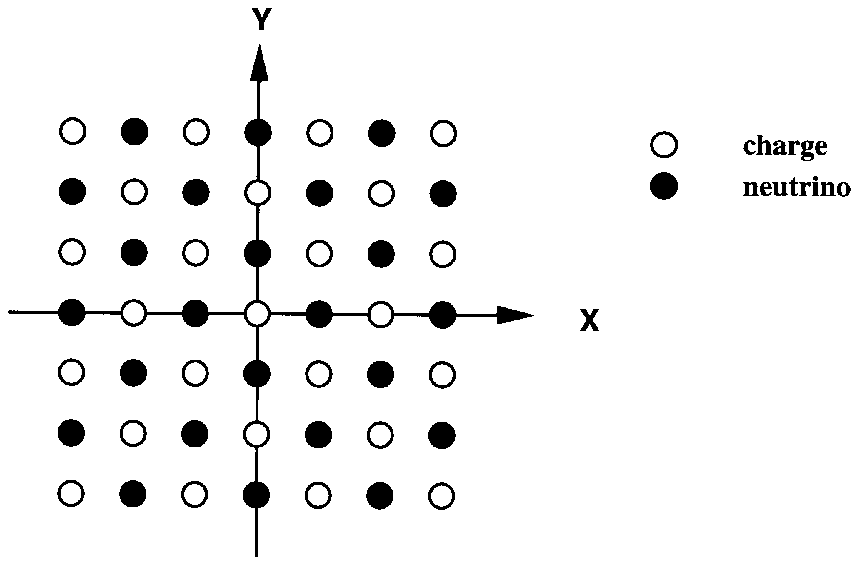}
\vspace{-0.2cm}
\begin{quote}
Fig.1. Horizontal or vertical section through the central part of\\ 
\indent\hspace{1.1cm} the electron lattice.
\end{quote}
\end{figure}

   The \emph{rest mass of the muon}  has been explained similarly with an 
oscillating lattice of muon and electron neutrinos [13]. 
 We found that m($\mu^\pm)$/m(e$^\pm$) is\\  
$\cong$ 3/2$\alpha_f$  = 205.55,  nearly equal to the actual mass ratio
 206.768,  in agreement with what Nambu [16] found empirically.
 The heavy weight of the muon is primarily a consequence
of the heavy weight of the N$^\prime$/4 muon neutrinos in the muon lattice.
 The mass of the muon neutrino is related to the mass of the electron
 neutrino through m($\nu_e)$ = $\alpha_f$m($\nu_\mu$), Eq.(39) of [13]. 

   In order to confirm the \emph{validity} of our preceding explanation of the 
mass of the electron we must show that the sum of the charges of the electric 
oscillations in the interior of the electron is equal to the elementary electric 
charge of the electron.   
We recall that Fourier analysis requires that, after pair production,
there must be a continuum of frequencies in the electron and positron.
With h$\nu_0\alpha_f$ = e$^2$/\emph{a} from Eq.(6) follows 
from Eq.(5) that the oscillation energy in e$^\pm$ is the sum of 
2$\times$(N$^\prime$/4 + 1) $\cong$ N/2  electric oscillations

\begin{equation} \mathrm{E}_\nu(\mathrm{e}^\pm) =  
\frac{\mathrm{N}}{2} \cdot \frac{\mathrm{e^2}}{\emph{a}}\cdot 
\frac{f(T)}{2\pi}\,
\int\limits_{-\pi}^{\pi}\,\phi\,\mathrm{d}\phi\,, \end{equation}
with f(T) = 1/(e$^{h\nu/kT} \mathrm{-}$ 1) = 1/1.305$\cdot10^{13}$
from p.17 in [13]. Inserting the values for N, f(T) and \emph{a} we find
that E$_\nu$(e$^\pm$) = 0.968\,m(e$^\pm$)c$^2$/2. The discrepancy
between m(e$^\pm$)c$^2$/2 and E$_\nu$(e$^\pm$) so calculated
must originate from the uncertainty of the parameters N, f(T) and \emph{a}
in Eq.(13). We note that it follows from the factor e$^2$/\emph{a} in
 Eq.(13) that the oscillation energy is the same for electrons and positrons, 
as it must be.

   We replace the integral divided by 2$\pi$ in Eq.(13), which has the value $\pi$/2,
 by the sum $\Sigma\,\phi_k\Delta\,\phi$, where k is an integer number with the
 maximal value k$_m$ = (N/4)$^{1/3}$. $\phi_k$ is equal to k$\pi$/k$_m$ 
and we have

\begin{displaymath}  \Sigma\,\phi_k\,\Delta\phi = \sum_{k=1}^{k_m}\,\frac{k\pi}{k_m}\cdot\frac{1}{k_m}
= \frac{ k_m(k_m + 1)\pi}{2\,k_m^2} \cong \frac{\pi}{2}\,,
\end{displaymath}
\noindent as it must be. The energy in the individual electric oscillation with
 index k is then 

\begin{equation} \Delta\mathrm{E}_\nu(k) = \phi_k\,\Delta\phi = k\pi/k_m^2\,.
\end{equation}

      Suppose that the energy of the electric oscillations is correctly 
described by the self-energy of an electrical charge
 
\begin{equation}\mathrm{U} = 1/2\,\cdot\,\mathrm{e}^2/\mathrm{r}\,.
\end{equation}
The self-energy of the elementary electrical charge is normally used to 
determine the mass of the electron from its charge, here we use Eq.(15) 
the other way around, we determine the charge from the energy  
in the oscillations. 

   The charge of the electron is contained in the electric oscillations. 
That means that \emph{the electric charge is not concentrated in a
point} but is distributed over N/4 = O($10^9)$ charge elements
 Q$_k$. \emph{The charge elements are distributed in a cubic lattice} and the 
resulting electric field is cubic, not spherical. For distances large as compared 
to the sidelength of the cube, (which is O($10^{-13}$)\,cm), say at the first
Bohr radius which is on the order of $10^{-8}$\,cm, the deviation of the cubic 
field from the spherical field will be reduced by about $10^{-10}$.   
 The charge in all electric oscillations is  
\begin{equation} \mathrm{Q} = \sum_{k}\,\mathrm{Q}_\mathrm{k}\,. \end{equation}

   Setting the radius r in the formula for the self-energy  equal to 2\,\emph{a} we find, 
with Eqs.(13,14,15), that the charge in the individual electric oscillations is 

\begin{equation} \mathrm{Q_k} = \pm\,\sqrt{2\pi\,N\,e^2f(T)/k_m^2}\,\cdot\,\sqrt{k}\,.
\end{equation}  

\noindent
and with k$_m$ = 1/2\,(N/4)$^{1/3}$ = 447 and
\begin{displaymath}  \sum_{k=1}^{k_m}\,\sqrt{k} = 6310.8\, \end{displaymath}

\noindent follows, after we have doubled the sum over $\sqrt{k}$, because for 
each index k there is a second oscillation on the negative axis of $\phi$, that

\begin{equation} \mathrm{Q} = \Sigma\,\mathrm{Q_k} = 
\pm\,5.027\cdot10^{-10}\,\,\mathrm{esu}\,,\end{equation}

\noindent whereas the elementary electrical charge is e = 
$\pm$\,4.803\,$\cdot\,10^{-10}$\,esu.  That means that our theoretical charge of the 
electron is 1.047 times the elementary electrical charge. Within the uncertainty
 of the parameters the theoretical charge of the electron agrees with
the experimental charge e. We have confirmed that it follows from our 
explanation of the mass of the electron that the electron has, within a 
5\% error, the correct electrical charge.     

   Each element of the charge distribution is surrounded in the horizontal
plane by four electron neutrinos as in Fig.(1), and in vertical direction by an
 electron neutrino above and also below the element. The electron neutrinos 
hold the charge elements in place.  We must assume that the charge 
elements are bound to the neutrinos by the weak nuclear force. The 
weak nuclear force plays here a role similar to its role in holding, for example,  
the $\pi^\pm$ or $\mu^\pm$ lattice together. It 
is not possibe, in the absence of a definitive explanation
 of the neutrinos, to give a theoretical explanation for the electro-weak 
interaction between the electric oscillations and the neutrinos. 
However, the presence of the
 range \emph{a} of the weak nuclear force in e$^2$/\emph{a} is a sign that 
the weak force is involved in the electric oscillations. The attraction of the 
charge elements by the
neutrinos overcomes the Coulomb repulsion of the charge elements.
 The weak nuclear force is the missing non-electromagnetic force or the 
Poincar\'{e} stress which holds the elementary electric charge together. 
The same considerations 
apply for the positive electric charge of the positron, only that then the 
electric oscillations are all of the positive sign and that they are 
bound to  anti-electron neutrinos.

   Finally we learn that  Eq.(13) precludes the possibility that the charge of 
the electron sits only on its surface. The number N in Eq.(13) would then be 
on the order of $10^6$, whereas N must be on the order of $10^9$ so that
E$_\nu$(e$^\pm$) can be m($\mathrm{e}^\pm)$c$^2$/2 as is necessary. 
In other words, the charge of the electron must be distributed throughout the 
interior of the electron, as we assumed.  
 
   Summing up: The rest mass of the electron and positron originates from 
the sum of the rest masses of N$^\prime$/4 electron neutrinos or anti-electron 
neutrinos in cubic lattices plus the mass in the energy of N$^\prime$/2 electric
 oscillations in the neutrino lattices. That means that neither the electron nor the 
positron are point particles. The electric oscillations are attached to the 
neutrinos by the weak nuclear force. The sum of the charge elements
 of the electric oscillations 
accounts for the elementary charge of the electron, respectively positron.

\section{The spin and magnetic moment \\ of the electron} 

    The model of the electron we have proposed in the preceding
 chapter has, in order to be valid, to pass a crucial test; the model has 
to explain satisfactorily the spin and the magnetic moment of the
 electron. When Uhlenbeck and Goudsmit [8] (U\&G) discovered
 the existence of the spin of the electron they also proposed that the 
electron has a magnetic moment with a value equal to Bohr's 
magnetic moment $\mu_B$ = e$\hbar$/2m$(\mathrm{e}^\pm)$c. Bohr's 
magnetic moment results from the motion of an electron on a circular  
orbit around a proton. The magnetic moment of the electron postulated
by U\&G has been confirmed experimentally, but has been corrected by 
  about  0.11\% for the so-called anomalous magnetic moment. 
If one tries to explain the magnetic moment of the electron
  with an electric charge moving on a circular orbit around the 
particle center, analogous to the magnetic moment of hydrogen, one ends 
up with velocities larger than the velocity of light, which cannot be, as 
already noted by U\&G. It remains to be explained how the magnetic
 moment of the electron comes about.

   We will have to explain the spin of the electron first. The spin, or 
the intrinsic angular momentum of a particle is, of course, the sum of the 
angular momentum vectors of all components of the particle. In the  
 electron these are the neutrinos and the electric oscillations. Each 
neutrino has spin 1/2 and in order for the electron to have 
s = 1/2 all, or all but one, of the spin vectors of the neutrinos in their 
lattice must cancel.
If the neutrinos are in a simple cubic lattice as in Fig.(1) and the
 center particle of the lattice is not  a neutrino, as in Fig.(1), the 
spin vectors of all neutrinos in the lattice cancel, $\Sigma\,j(n_i)$ = 0, 
provided that the spin vectors of the
electron neutrinos of the lattice point in opposite direction at their
 mirror points in the lattice. Otherwise the spin vectors of the neutrinos 
would add up and make a very large angular momentum. We 
follow here the procedure we used in [17] to explain the spin
of the muons. The spin vectors of all electron neutrinos in the electron 
cancel just as the spin vectors of all muon and electron neutrinos
 in the muons cancel because there is
 a neutrino vacancy at the center of their lattices, (Fig.(1) of [17]).

   We will now see whether the electric oscillations in the electron 
contribute to its angular momentum. 
As we said in context with Eq.(7) there must be two times 
as many electric oscillations in the electron lattice than there are 
neutrinos. The oscillation pairs can either be the two oscillations in a 
standing wave or they can be  two circular oscillations. Both the standing 
waves and the circular oscillations are non-progressive and can be part of
 the \emph{rest mass} of a particle. We will now assume that the electric
 oscillations are circular.  Circular oscillations have an
 angular momentum $\vec{j} = m\,\vec{r}\times\vec{v}$. And, as in the case 
of the spin vectors of the neutrinos, all or all but one of the O$(10^9)$ 
angular momentum vectors of the electric oscillations must cancel in order 
for the electron to have spin 1/2. As in [13] we will describe the 
superposition of the two circular oscillations by

\begin{equation} x(t) = exp[i\omega t] + exp[-\,i(\omega t + 
\pi)]\,,\end{equation}

\begin{equation} y(t) = exp[i(\omega t + \pi/2)] + exp[-\,i(\omega t + 
                 3\pi/2)]\,\,,
\end{equation} 
\noindent
that means by the superposition of a circular oscillation with the 
frequency $\omega$ and a second circular oscillation with the frequency
 $\mathrm{-}\,\omega$. The latter oscillation is shifted in phase by 
$\pi$. Negative frequencies are permitted solutions of the equations 
of motion in a cubic lattice, Eqs.(7,13) of [13]. 
As is well-known oscillating electric charges should emit radiation.
However, this rule does already not hold in the hydrogen atom, so we
will assume that the rule does not hold in the electron either.

   In circular oscillations the kinetic energy is always equal to the potential 
energy  and the sum of both is the total energy. From
\begin{equation} \mathrm{E}_{pot} + \mathrm{E}_{kin} = 2\,\mathrm{E}_{kin}
=\mathrm{E}_{tot} \end{equation}
follows with E$_{kin}$ = $\Theta\,\omega^2$/2 and E$_{tot}$ = $\hbar\omega$
that 2\,E$_{kin}$  = $\Theta\,\omega^2$ = $\hbar\omega$. 
 ${\Theta}$ is the moment of inertia. When we superpose the two
 circular oscillations with $\omega$ and $\mathrm{-}\,\omega$ of Eqs.(19,20)
 we have  
\begin{equation} 2\times2\,\mathrm{E}_{kin} = 2\,\Theta\,\omega^2 =
 \hbar\omega\,,\end{equation}
from which follows that the angular momentum is
\begin{equation} j = \Theta\,\omega = \hbar/2\,. \end{equation} 
That means that each of the O$(10^9$) pairs of superposed circular 
oscillations has an angular momentum  $\hbar$/2.

   The circulation of the oscillation pairs in Eqs.(19,20) is opposite for 
all $\omega$ of opposite sign. It follows from the equation for the 
displacements u$_n$  of the lattice points
\begin{equation} u_n = Ae^{i(\omega\,t\, +\, n\phi)}\,,
\end{equation}
\noindent
(Eq.(5) in [13]) that the velocities of the lattice points are given by
\begin{equation} v_n = \dot{u}_n = i\,\omega_n\,u_n\,. \end{equation}
The sign of $\omega_n$ changes with the sign of $\phi$ because the 
frequencies are given by Eq.(13) of [13], that means by  
\begin{equation} \omega_n = \pm\,\omega_0\,[\,\phi_n + \phi_0\,]\,.
\end{equation}
Consequently the circulation of the electric oscillations is opposite to 
the circulation at the
mirror points in the lattice and the angular momentum vectors cancel,
but for the angular momentum vector of the electric oscillation at 
the\,\emph{ center of the lattice}. The center circular oscillation has, as all 
other electric oscillations, the angular momentum  $\hbar$/2 as Eq.(23) 
says. The angular momentum of the entire electron lattice is therefore
\begin{equation} j(\mathrm{e}^\pm) = \sum\,j(n_i) + \sum\,j(el_i) = j(el_0) 
=\hbar/2\,,
\end{equation}
as it must be for spin s = 1/2. The explanation of the spin of the electron given
here follows the explanation of the spin of the baryons  in [13], as well
as the explanation of the absence of spin in the mesons. A valid explanation 
of the spin must be applicable to all particles, in particular to the electron, the 
prototype of a particle with spin. 

   We will now turn to the magnetic moment of the electron which is known  
with extraordinary accuracy, $\mu(\mathrm{e}^\pm)$ = 
1.001\,159\,652\,187\,$\mu_B$,  according
 to the Review of Particle Physics [14], with $\mu_B$ being the Bohr 
magneton. The decimals after 1.00\,$\mu_B$ 
are caused by the anomalous magnetic moment which we will not consider. 
As is well-known the magnetic dipole moment of a particle with spin is, 
in Gaussian units, given by
\begin{equation} \vec{\mu} = g\,\frac{e\hbar}{2mc}\,\vec{s}\,, \end{equation}
where g is the dimensionless Land\'{e} factor, m the rest mass of the 
particle and $\vec{s}$ the spin vector.
 The g-factor has been introduced in order to bring the magnetic moment
 of the electron into agreement with the experimental facts. As U\&G
 postulated and as has been confirmed experimentally the 
g-factor of the electron is 2. With the spin s = 1/2 of the electron the 
magnetic dipole moment of the electron is then
\begin{equation} \mu(\mathrm{e}^\pm) = 
\mathrm{e}\hbar/2\mathrm{m}(\mathrm{e}^\pm)c\,, \end{equation}
or one Bohr magneton in agreement with the experiments, neglecting the
 anomalous moment. For a structureless
point particle Dirac [9] has explained why g = 2 for the electron. However 
we consider here an electron with a finite size and which is at rest, which 
means that the velocity of the center of mass is zero. When it is at
 rest the electron has still its magnetic moment. Dirac's theory does 
therefore not apply here.

   The only part of Eq.(28) that can be changed in order to explain the 
g-factor of an electron with structure is the ratio e/m which deals with the 
spatial distribution of charge and mass. In the classical electron models
 the mass originates from the charge.
 However that is not necessarily always so. If part of the mass of the 
electron is non-electrodynamic and the non-electrodynamic part of the mass
 does not contribute to the magnetic moment of the electron, which to all 
that we know is true for neutrinos, then the ratio e/m in Eq.(28) is not
 e/m($\mathrm{e}^\pm$)  in the case of the electron. The elementary charge 
e certainly remains unchanged, but  e/m depends on what fraction of the 
mass is of electrodynamic origin and what fraction of m is 
 non-electrodynamic, just as the mass of a current loop does not contribute
to the magnetic moment of the loop. From the very accurately known values
of $\alpha_f$, m($\pi^\pm$)c$^2$ and m(e$^\pm$)c$^2$ and from Eq.(9)
for the energy in the electric oscillations in the electron E$_\nu$(e$^\pm$) = 
$\alpha_f$/2\,$\cdot$\,m($\pi^\pm$)c$^2$/2 = 
0.996570\,m(e$^\pm$)c$^2$/2 follows   
that very nearly one half of the mass of the  electron is of electric origin, 
whereas the other half  of m($\mathrm{e}^\pm)$  is made of neutrinos
which do not contribute to the magnetic moment. That means that in the  
electron the mass in e/m is practically  m($\mathrm{e}^\pm$)/2. The 
magnetic moment of the electron is then 
\begin{equation}
\vec{\mu}_e = g \frac{e\hbar}{2m(e^\pm)/2\cdot c}\vec{s}\,,
 \end{equation}
and with s = 1/2 we have $\mu(\mathrm{e}^\pm)$ = 
g\,e$\hbar$/2m($\mathrm{e}^\pm$)c. Because of 
Eq.(29) the g-factor must be equal to one and is unnecessary. In other words,
 if the electron is composed of the neutrino lattice and the electric oscillations 
as we have suggested, then the electron has the correct magnetic moment 
$\mu_e$ = e$\hbar/2\mathrm{m(e^\pm)c}$,  if exactly 1/2 of the electron
 mass consists of neutrinos.

   The preceding explanation of the magnetic moment of the electron has
 to pass a critical test, namely it has to be shown that the same
 considerations lead to a correct explanation of the magnetic moment of 
the muon $\mu_\mu$ = e$\hbar$/2m($\mu^\pm$)c,  which is about 1/200th 
of the magnetic  moment of the electron but is known with nearly the same
accuracy as $\mu_e$.  Both magnetic moments are related  
through the equation
\begin{equation} \frac{\mu_\mu}{\mu_e} = 
\frac{\mathrm{m(e}^\pm)}{\mathrm{m}(\mu^\pm)} = \frac{1}{206.768}\,,
 \end{equation}
as follows from Eq.(28) applied to the electron and muon.
This equation agrees with the experimental results to the sixth decimal.
The muon has, as the electron, an anomalous magnetic moment of about
0.11\,\% $\mu_\mu$, which is too small to be considered here.

   In the standing wave model [13] the muons consist of a lattice of 
N$^\prime$/4 muon neutrinos $\nu_\mu$, respectively anti-muon 
neutrinos $\bar{\nu}_\mu$, of N$^\prime$/4 electron neutrinos and 
the same number of anti-electron neutrinos plus an elementary electric 
charge. For the explanation of the magnetic moment of the muon we 
follow the same reasoning we have used for the explanation of the 
magnetic moment of the electron. We say that m($\mu^\pm)$  consists 
of two parts, one part which causes the magnetic moment and another 
part which does not contribute to the magnetic moment.  The part
of m($\mu^\pm$) which causes the magnetic moment must contain  
 circular electric oscillations without which there would 
be no magnetic moment. It becomes immediately clear from the small
mass of the electron neutrinos and from Eq.(5) for the energy of the 
 electric oscillations in the electron that $\Sigma$\,m($\nu_e$) and 
 E$_\nu$(e$^\pm$) are too small, as compared to the energy in the 
rest  masses of all neutrinos in the muons, to make up m($\mu^\pm)$/2.
However, the oscillations in the $\mu^\pm$\,mesons do not follow 
Eq.(5) for the oscillation energy in the electron, but rather Eq.(4) for
the oscillation energy in the muons. Both differ by the factor $\alpha_f$ 
in Eq.(5). But even when the oscillation energy in the muons as given 
by Eq.(4) is considered, the energy of the electric oscillations in the 
muons would be only E$_\nu(\mu^\pm)$/4 = 16.955\,MeV, if the 
electric oscillations are attached to N/4 electron neutrinos, as is the 
case in the electron.

   It appears to be necessary to consider the case that the electric 
oscillations in the $\mu^\pm$\,mesons are attached to \emph{all} 
neutrinos of the electron neutrino type in the $\mu^\pm$ lattice, 
regardless whether they are 
electron neutrinos or anti-electron neutrinos. That would mean that 
the electric charge is distributed uniformly in the $\mu^\pm$ lattice.
There are, as has been shown in the paragraph below Eq.(31) of [13],
3/4$\cdot$N neutrinos of the electron neutrino type in the muons, of
which N/4 neutrinos originate from the charge e$^\pm$ carried by 
$\mu^\pm$. If the electric oscillations are attached to 3/4$\cdot$N
electron neutrinos, regardless of their type, then the energy in all 
electric oscilllations or the energy in the electric charge is, with Eq.(8) 
and E$_\nu(\mu^\pm$) = E$_\nu(\pi^\pm$) = 67.82\,MeV from 
Eq.(31) in [13], as well as with m($\mu^\pm$)c$^2$ = 105.6583\,MeV, 
given by
\vspace{1cm}
\begin{eqnarray}
\lefteqn{3/4\cdot\mathrm{E}_\nu(\mu^\pm) = 
3/4\cdot 67.82\,\mathrm{MeV} = 50.865\,\mathrm{MeV}}\nonumber\\&
 = & 0.4814\,\mathrm{m}(\mu^\pm)\mathrm{c}^2 \cong 
1/2\cdot \mathrm{m}(\mu^\pm)\mathrm{c}^2\,. \end{eqnarray}
In other words, the energy in the electric oscillations or the electric
charge makes up, in a good approximation, 1/2 of the mass of the muons.
The other half of the rest mass of the muons consists of the sum of the 
rest masses of the neutrinos in the muon lattice plus the oscillation energy
of the muon neutrinos, neither of which contributes to the magnetic moment.
It is

\begin{equation} 1/4\cdot\mathrm{E}_\nu(\mu^\pm) +
\mathrm{N}/4\cdot\mathrm{m}(\nu_\mu)\mathrm{c}^2 +
3/4\cdot\mathrm{Nm}(\nu_e)\mathrm{c}^2 \\
= 53.347\,\mathrm{MeV} = 0.50490\,\mathrm{m}(\mu^\pm)\mathrm{c}^2\,.
\end{equation} 
The theoretical total energy in the rest mass of the muons is then 
E(m($\mu^\pm$)) = 0.9863\,m($\mu^\pm$)c$^2$(exp).

   In simple terms, if E$_\nu(\pi^\pm)$ = E$_\nu(\mu^\pm$) =
1/2$\cdot$m($\pi^\pm$), not 0.486\,m($\pi^\pm$) as in Eq.(27)
of [13], then it follows from 3/4$\cdot$E$_\nu(\mu^\pm$) = 
3/8$\cdot$m($\pi^\pm$) and from the neutral part of the muon mass in Eq.(33),
which is likewise $\approx$3/8$\cdot$m($\pi^\pm$), that the rest mass of the 
muons is m($\mu^\pm$) $\cong$ 3/8$\cdot$m($\pi^\pm$) + 3/8$\cdot$m($\pi^\pm$) 
= 3/4$\cdot$ m($\pi^\pm$), as it must be in a first approximation, 
whereas the actual m($\mu^\pm$) is 
1.00937$\cdot$3/4$\cdot$m($\pi^\pm$). That means that in a good
approximation the charged part of the rest mass of the muons is 1/2 
of the mass of the muons.

   If the charged part of the muon mass as expressed by Eq.(32) makes up
1/2 of the mass of the muons and if the other part of the muon mass does
 not contribute to the magnetic moment, then the magnetic moment of the 
muon is given by  
\begin{equation} \vec{\mu}_\mu = \frac{\mathrm{e}\hbar}
{2\mathrm{m}(\mu^\pm)/2\cdot\mathrm{c}}\cdot\vec{s}\,.  \end{equation}
With s = 1/2 we have $\mu_\mu$ = e$\hbar$/2m$(\mu^\pm)$c as it must 
be, without the artificial g-factor.

\section*{Conclusions}

   One hundred years of sophisticated theoretical work have made it abundantly 
clear that the electron is not a purely electromagnetic particle. There must be 
something else in the electron but electric charge. It is equally clear from the 
most advanced scattering experiments that the  ``something else"  in the 
electron must be non-interacting, otherwise it could not  be that  we find that 
the radius of the electron must be smaller than $10^{-16}$\,cm. The only 
non-interacting matter we know of with certainty are the neutrinos. So it seems 
to be natural to ask whether neutrinos are not part of the electron. Actually
we have not introduced the neutrinos in an axiomatic manner but rather as a 
consequence of our standing wave model of the stable mesons, baryons
and $\mu$-mesons. It follows necessarily from this model that after the 
decay of the $\mu^-$\,meson there must be electron neutrinos in the emitted 
electron, and that they make up one half of the mass of the electron. The other
half of the energy in the electron originates from the energy of electric 
oscillations. The theoretical rest mass 
 of the electron  agrees, within 1\% accuracy, with the experimental
 value of m(e$^\pm$). We have learned that the charge of the electron
 is not concentrated in a single point, but rather is distributed over O(10$^9$)
elements which are held together with the neutrinos by the weak nuclear force. 
The sum of the charges in the electric oscillations is, within the accuracy of the
 parameters, equal to the elementary electrical charge of the electron.
  From the explanation of the mass and charge of the electron follows, 
 as it must be, the correct spin and magnetic moment of the electron, the
other two fundamental features of the electron. With a cubic lattice of anti-electron
neutrinos we also arrive with the same considerations as above at the correct 
mass, charge, spin and magnetic moment of the positron.

\section*{Acknowledgements}

Contributions of Professor J. Zierep and of Dr. T. Koschmieder are gratefully
 acknowledged.

\section*{References}

\noindent
[1] Thomson, J.J. Phil.Mag. {\bfseries44},293 (1897).

\smallskip
\noindent
[2] Lorentz, H.A. \emph{Enzykl.Math.Wiss.} Vol.{\bfseries5},188 (1903).

\noindent
[3] Poincar\'{e}, H. Compt.Rend. {\bfseries 140},1504 (1905). Translated 
in:\\
\indent
 Logunov, A.A. \emph{On The Articles by Henri Poincare\\ 
\indent
``On The Dynamics of the Electron"}, Dubna JINR (2001).

\smallskip
\noindent
[4] Ehrenfest, P. Ann.Phys. {\bfseries24},204 (1907).

\smallskip
\noindent
[5] Einstein, A. Sitzungsber.Preuss.Akad.Wiss. {\bfseries20},349 (1919).

\smallskip
\noindent
[6] Pauli, W. \emph{Relativit\"{a}tstheorie}, B.G. Teubner (1921). 
Translated in:\\
\indent  Theory of Relativity, Pergamon Press (1958).

\smallskip
\noindent
[7] Poincar\'{e}, H. Rend.Circ.Mat.Palermo {\bfseries21},129 (1906).

\smallskip
\noindent
[8] Uhlenbeck, G.E. and Goudsmit, S. Naturwiss. {\bfseries13},953 (1925).

\smallskip
\noindent
[9] Dirac, P.A.M. Proc.Roy.Soc.London A{\bfseries117},610 (1928).

\smallskip
\noindent
[10] Gottfried, K. and Weisskopf, V.F. \emph{Concepts of Particle 
Physics},\\
\indent
 \,\,Vol.1,\,p.38. Oxford University Press (1984).

\smallskip
\noindent
[11] Schr\"{o}dinger, E. Sitzungsber.Preuss.Akad.Wiss. {\bfseries24},418 
(1930).

\smallskip
\noindent
[12] Mac Gregor, M.H. \emph{The Enigmatic Electron}, Kluwer (1992).

\smallskip
\noindent
[13] Koschmieder, E.L. http://arXiv.org/phys/0602037 (2006).

\smallskip
\noindent
[14] Eidelman, S. et al. Phys.Lett.B {\bfseries 592},1 (2004).

\smallskip
\noindent
[15] Born, M. and v.\,Karman, Th. Phys.Z. {\bfseries13},297 (1912). 

\smallskip
\noindent
[16] Nambu, Y. Prog.Th.Phys. {\bfseries7},595 (1952).

\smallskip
\noindent
[17] Koschmieder, E.L. http://arXiv.org/physics/0308069 (2003),\\
\indent\,\, \emph{Muons: New Research}, Nova (2005).

\end{document}